\documentclass[11pt]{article} 
\input epsf.sty

\def\square{\rule{2mm}{2mm}}

\def\squarebox#1{\hbox to #1{\hfill\vbox to #1{\vfill}}}

\newcommand{\tensor}{\otimes}

\newcommand\meet\wedge
\newcommand\implies\Rightarrow

\def\ccc{{\mathchoice {\setbox0=\hbox{$\displaystyle\rm
C$}\hbox{\hbox
to0pt{\kern0.4\wd0\vrule height0.9\ht0\hss}\box0}}
{\setbox0=\hbox{$\textstyle\rm C$}\hbox{\hbox
to0pt{\kern0.4\wd0\vrule height0.9\ht0\hss}\box0}}
{\setbox0=\hbox{$\scriptstyle\rm C$}\hbox{\hbox
to0pt{\kern0.4\wd0\vrule height0.9\ht0\hss}\box0}}
{\setbox0=\hbox{$\scriptscriptstyle\rm C$}\hbox{\hbox
to0pt{\kern0.4\wd0\vrule height0.9\ht0\hss}\box0}}}}

\newcommand{\complex}{{\mathcal C}}
\newcommand{\reals}{{\hbox{\sf I\kern-.14em\hbox{R}}}}

\newcommand{\size}[1]{\left|#1\right|}

\newcommand{\ket}[1]{|#1\rangle}
\newcommand{\bra}[1]{\langle #1|}
\newcommand{\braket}[2]{\langle #1 | #2\rangle}
\newcommand{\ketbra}[2]{\ket{#1}\!\bra{#2}}
\newcommand{\density}[1]{\ketbra{#1}{#1}}

\newcommand{\eqdef}{\stackrel{\rm def}{=}}

\newcommand{\ignore}[1]{}

\newcommand{\zee}{{\mathcal Z}}
\newcommand{\ee}{{\mathrm e}}
\newcommand{\tpsi}{{\tilde\Psi}}
\newcommand{\teff}{{\tilde f}}
\newcommand{\aai}{{\mathrm i}}

\newcommand{\lft}{{\mathrm L}}
\newcommand{\rght}{{\mathrm R}}
\newcommand{\slow}{{\mathrm{slow}}}
\newcommand{\fast}{{\mathrm{fast}}}

\newcommand{\tpee}{{\tilde{P}}}

\begin{document}

\title{\Large {\bf Quantum Walk on the Line} \\
                (Extended Abstract)
}

\author{
     Ashwin Nayak
     \thanks{Supported by a joint DIMACS-AT{\&}T Post-Doctoral
             Fellowship and NSF grant EIA~00-80234.} \\
     DIMACS Center \\
     Rutgers University \\
     P.O.\ Box~1179 \\
     Piscataway, NJ~08855 \\
     {\tt nayak@dimacs.rutgers.edu}
     \and
     Ashvin Vishwanath
     \thanks{Supported by a C.E.\ Procter Fellowship.} \\
     Joseph Henry Laboratories \\
     Department of Physics \\
     Princeton University \\
     Princeton, NJ~08544 \\
     {\tt ashvinv@princeton.edu}
}

\date{}

\maketitle

\begin{abstract}
Motivated by the immense success of random walk and Markov
chain methods in the design of classical
algorithms, we consider {\em quantum\/} walks on graphs.
We analyse in detail the behaviour of unbiased quantum walk on the line,
with the example of a typical walk, the ``Hadamard walk''. We show that
after~$t$ time steps, the probability distribution on the line induced 
by the Hadamard walk is almost uniformly distributed over the 
interval~$[-t/\sqrt{2},\;t/\sqrt{2}]$. This implies that the same walk
defined on the circle mixes in {\em linear\/} time.
This is in direct contrast with the quadratic mixing time for the
corresponding classical walk.
We conclude by indicating
how our techniques may be applied to more general graphs.
\end{abstract}

\section{Introduction}
\label{sec-intro}


Random walks on graphs have found many applications in computer science,
including randomised algorithms for
2-Satisfiability, Graph Connectivity and probability amplification
(see, e.g.,~\cite{MotwaniR95}). Recently,
Sch\"oning~\cite{Schoning99} discovered a
random walk based algorithm similar to that of~\cite{Papadimitriou91}
that gives an elegant (and the most efficient
known) solution to 3-Satisfiability.
In general, Markov chain simulation
has emerged as a powerful algorithmic tool and has had
a profound impact on random sampling and approximate
counting~\cite{JerrumS96}. Notable among its numerous applications are
estimating the volume of convex bodies~\cite{DyerFK91}\footnote{
See~\cite{LovaszK99} for the latest progress on this problem.} and
approximating the permanent~\cite{JerrumS89}. A few months ago,
Jerrum, Sinclair and Vigoda~\cite{JerrumSV00} used this approach to solve
the long standing open problem of approximating the permanent in the
general case.

In the spirit of developing similar techniques for quantum algorithms,
we consider {\em quantum\/} walk on graphs. To date, few general
techniques are known for developing and analysing quantum algorithms:
{\em Fourier sampling}, which is typified by the seminal work of
Simon~\cite{Simon97} and Shor~\cite{Shor97}, and {\em amplitude
amplification}, which originated in the seminal 
work of Grover~\cite{Grover96}.
Barring applications of these techniques,
the search for new quantum algorithms has primarily been {\it ad
hoc}. We believe that studying quantum walk on graphs is a step
towards providing a systematic way of speeding up classical
algorithms based on random walk.

We begin by considering the quantum walk in one dimension.
Recall that a classical particle doing a random walk on the integer
lattice chooses, at every time step, a random direction (``left'' or
``right'') to move in, and moves to the site adjacent to it in that
direction. In direct analogy, one may na\"ively try to define
quantum walk on the line as follows: at every time step, the particle
moves, {\em in superposition}, both left {\em and\/} right with
equal amplitudes (perhaps with a relative phase difference).
However, such a walk
is physically impossible, since the global process is non-unitary.
It is also easy to verify that
the only possible homogeneous (i.e., translationally invariant)
unitary processes on the line
involving transitions between adjacent lattice sites
are the left and right shift operators
(up to an overall phase)~\cite{Meyer96}. This corresponds to the rather
uninteresting motion in a single direction. As explained below (and
also shown by~\cite{Meyer96}),
it is still possible to construct a unitary walk if the particle
has an extra degree of freedom that assists in its motion. 

\begin{figure}
\begin{center}
\epsfxsize=2.0in
\hspace{0in}
\input{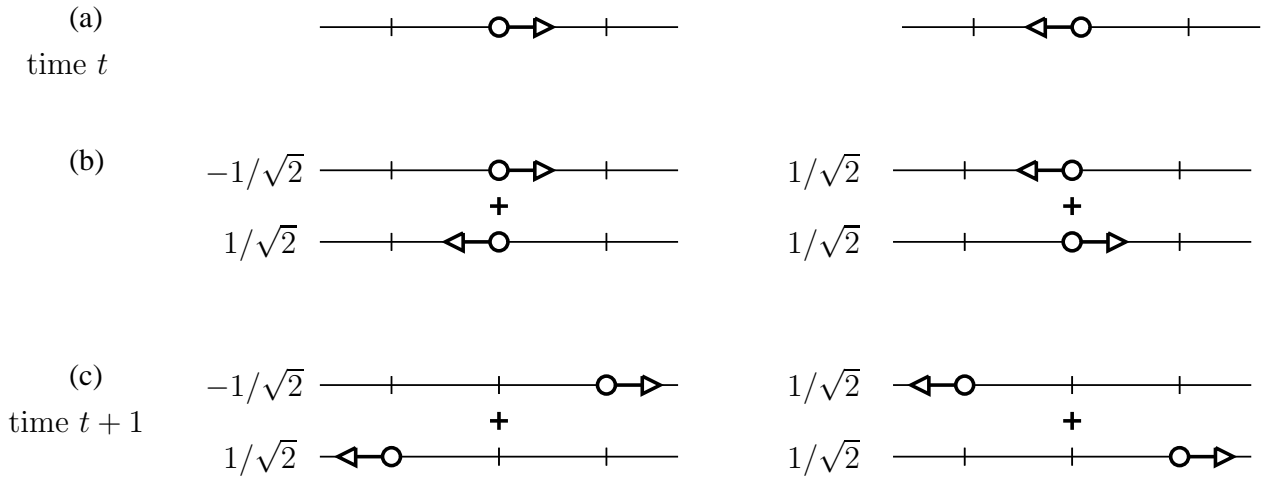}
\caption{\it The dynamics of the Hadamard walk. In (a) we begin at 
time~$t$ with a particle in chirality state ``right'' or ``left''.
The result of the Hadamard transformation is shown in (b),
the particle is now in an equal superposition of ``left'' and ``right''
chirality states (with the amplitudes indicated) and then moves
accordingly (c) to generate the state at time~$t+1$.}
\label{fig-hwalk}
\end{center}
\end{figure}

Consider a quantum particle that moves freely on the integer points on
the line, and has an additional degree of freedom, its {\em chirality},
which takes values ``left'' and ``right''. A walk on the line by such a
particle may be described as follows. At every time step, its chirality
undergoes a rotation (a unitary transformation in general) and then the
particle moves according to its final chirality state.
Figure~\ref{fig-hwalk} depicts this two-stage move in a quantum walk
where the chirality undergoes a Hadamard transformation. We call this
particular walk the ``Hadamard walk''.

In this paper, we analyse in detail the dynamics of the Hadamard
walk on the line. We derive the asymptotic form of the probability
induced on the line by observing the position of a particle doing the
walk and show that after~$t$ time steps, the distribution is almost
uniformly distributed over the interval~$[-t/\sqrt{2},\;t/\sqrt{2}]$.
This implies that the
analogously defined walk on the or
the circle {\em mixes\/} in {\em linear\/} time. This is in immediate 
contrast with the classical random walk, which mixes in {\em
quadratic\/} time.\footnote{In this paper,
by the $\delta$-{\em mixing time} of a walk we mean
the {\em first\/} time (independent of the initial state) at which
the distribution induced by the walk is~$\delta$-close to uniform in
total variation distance. The parameter~$\delta$ is understood to be a
constant less than~$1$ if it is not explicitly specified.
}

Next, we turn to general quantum walk on the line. Such a walk may
involve arbitrary transitions between adjacent lattice points on the
line (subject to the global unitarity constraint).
It turns out that any homogeneous, local
unitary process on the line
governing the motion of a particle 
may be recast as a unitary transformation of the chirality followed
by a left/right shift~\cite{NayakV00}. Moreover, if we are interested in
{\em unbiased\/} walk\footnote{Appendix~\ref{sec-general} has a formal
definition of this notion. Essentially, it means that the walk makes no
distinction between the two halves of the line given by its initial
position. It is not hard to see that in any other kind of
({\em biased\/}) unitary walk, the particle
moves in a single direction with probability~$1$ at every time step.},
we may assume that the particle moves one step to the left if it ends up
with chirality ``left'', and one step to the right if it ends up with
chirality ``right''. Thus, there is no loss in
generality in studying only such walk.

It can be shown~\cite{NayakV00} 
that if we are interested only
in the probability distribution induced on the line by such a walk, we
may restrict ourselves to the study of a family of walks specified by a
{\em single\/} parameter~$\theta \in [0,\pi]$.
We show that {\em every\/} walk in this family, except that for~$\theta =
0,\pi$, shares the gross features of the Hadamard walk. In
particular, all except the two singular walks mix in linear time.


One method for analysing quantum processes such as the walk we study is
the {\em path integral\/} approach. This method involves
explicitly computing the amplitude of being in a certain state as
the sum,
over all possible paths leading to that state, of the amplitude of
taking that path. The problem of evaluating, or even approximating, path
integrals to determine the time development of a wave function is well
known to be notoriously hard.
We circumvent this barrier by taking the {\em Sch\"odinger approach},
which takes advantage of the time and space homogeneity
of the quantum walk. The crucial observation is that because of its
translational invariance, the walk has a simple description in
Fourier space. The Fourier transform of the wave function is thus easily
analysed, and transformed back to the spatial domain. It is noteworthy
that this technique is standard in the analysis of
classical random walk~\cite{Diaconis88}.

A key advantage of the Sch\"odinger approach is that the resulting
description of the wave function,
in terms of integrals of the Fourier type,
is amenable to analysis in standard ways. 
There is a well-developed theory of the asymptotic expansion of
integrals~\cite{BenderO78}
which allows us to determine the leading behaviour of the
wave function in the large time limit. (This is exactly the case of
interest in the design of asymptotically efficient algorithms.)
We are thus able to derive the asymptotic functional form
of the corresponding probability distribution.

We also use the theory of asymptotic expansion of integrals in a
non-standard way
to deduce some properties of the probability distribution induced by
the walk. For example, we use this method to 
approximate the probability masses of each point on the line 
by an integral of the Fourier type. This
allows us to calculate the net probability of an interval, and the
different moments of the distribution very precisely.

Finally, we note that the asymptotic quantities we calculate
match very well with simulation results even for small times,
indicating that our results may apply to small quantum systems as well.

\subsection*{Related work}

Quantum dynamics similar to ours has previously been studied
by Meyer~\cite{Meyer96} in the context of quantum cellular automata. 
He singles out the one parameter family of unitary processes
mentioned above and derives a
closed form solution for the corresponding wave functions.
The solution is obtained by following the path integral approach and
allows him to express it in terms of the Gauss
hypergeometric function. His motivation in studying such dynamics is
however very different from ours, and he does not quantitatively
analyse the wave functions further.

In work more closely related to ours in its motivation, 
Nayak, Schulman and Vazirani~\cite{NayakSV97} have studied random sampling
on a quantum computer. They gave an efficient method
for generating superpositions for which the corresponding probability
distributions satisfy certain well-behavedness properties.
Using this for a distribution related by a Fourier transform,
they gave an efficient
quantum algorithm for sampling from the Gibbs distribution for the
Ising model. (Subsequently, Randall and Wilson~\cite{RandallW99}
discovered a {\em classical\/} polynomial time algorithm for this task
via a different route.)

Watrous~\cite{Watrous99a} has considered unitary processes based on
quantum walk
on regular graphs in the context of Undirected Graph Connectivity and
logarithmic-space computation. He shows that
it is possible to construct, in logarithmic space
with limited measurement, and with high probability,
a good approximation to the superposition
over all vertices connected to a specified vertex in the graph, using a
local unitary transform as a subroutine.

We have recently learnt that Aharonov, Ambainis, Kempe
and Vazirani~\cite{AharonovAKV00}
have independently studied quantum walks
on graphs and their mixing behaviour.
They observe that the probability
distribution induced by a unitary process on a finite dimensional
Hilbert space is essentially periodic and hence does not converge to a
stationary distribution. They argue that the notion of
mixing in this case is therefore more appropriately captured by the 
closeness to the uniform distribution, of
the {\em average\/} of the probability distribution
over all time steps.
They show that the quantum walk on the~$n$-circle
mixes in time~$O(n\log n)$ in this sense. They also show a lower
bound of~$1/d\Phi$ for this mixing time for general graphs,
where~$d$ is the maximum degree of a vertex in the graph
and~$\Phi$ is the {\em conductance}.

We have also learnt that Ambainis and Watrous~\cite{AmbainisW00}
have analysed the Hadamard walk on the unbounded line, and show that it
has linear mixing characteristics. They follow the 
approach of Meyer, and further 
analyse the path integrals so obtained to arrive
at the result.

\subsection*{Organisation of the paper}

The rest of the paper is organised as follows. We formally define and
analyse quantum walk on the line in Section~\ref{sec-walk}. Some
background on Fourier transforms required for this is summarised in
Appendix~\ref{sec-tech}. Details of some calculations required for the
section are provided in Appendix~\ref{App1}.
We then consider the behaviour of the walk
after large times and derive its asymptotic properties in
Section~\ref{sec-asympt}. This section is heavily based on the Method of
Stationary Phase described in Appendix~\ref{sec-tech}. The
approximations made in the section are justified in Appendix~\ref{App2}.
For lack of space, we are not able to provide the full details of the
analysis of the general quantum walk on the line. The technique involved
in the analysis is however the same as that for the Hadamard walk.
Appendix~\ref{sec-general} summarises the main conclusions about the
general case.

\section{One dimensional quantum walk}
\label{sec-walk}

\subsection{Formal description of the walk}

A particle doing a (classical) symmetric random walk on the line 
may be described as follows. It starts, say, at the origin, and 
at every time step, tosses a fair coin. Each of the two possible 
outcomes of the toss is associated with a distinct
direction, ``left'' or ``right''.
The particle moves one step in the direction resulting from the toss.

A quantum generalisation of this process involves a quantum particle on 
the line, but with an additional degree of freedom which we call the 
``chirality''. The chirality takes values ``left'' and ``right'', and
directs the motion of the particle. At any given 
time the particle may be in a superposition of ``left'' 
and ``right'' chirality states and is therefore described
by a two-component wave function. The dynamics of 
the walk that we consider is given by the following rules.
At each time step, the chirality undergoes a rotation (a unitary
transformation, in general), and the particle moves according to its
final chirality state. Therefore, if the particle ends up with chirality
``left'', it moves one step to the left, and if it ends up with
chirality ``right'', it moves one step to the right.

For concreteness, we begin by focusing on a quantum walk in which
the unitary transformation acting on the chirality state at each 
time step, is chosen to be the Hadamard transformation 
(the Fourier transform over~${\mathcal Z}_2$): 
\begin{eqnarray*}
\ket{\lft} & \mapsto & \frac{1}{\sqrt{2}} (\ket{\lft} + \ket{\rght}) \\
\ket{\rght} & \mapsto & \frac{1}{\sqrt{2}} (\ket{\lft} - \ket{\rght}).
\end{eqnarray*}
Here L and R refer to the ``right' and ``left'' chirality states.
The resulting quantum walk, which we will refer to as the ``Hadamard
walk'' is depicted in Figure~\ref{fig-hwalk}.

To study the properties of the walk defined above,
we consider the wave function describing the position of the
particle and analyse how it evolves with time.
Let~$\Psi(n,t)=\left( \begin{array}{c} \psi_{\lft}(n,t) \\ \psi_{\rght}(n,t) \end{array} \right) $ be the two component vector of amplitudes of the
particle being at point~$n$ at time~$t$, with the chirality being left (upper component) or right (lower component).
The dynamics for ~$\Psi$ is then given by the following transformation (cf.\
Figure~\ref{fig-hwalk}):
\begin{eqnarray*}
\Psi(n,t+1) & = & \left[  \begin{array}{cc}
                          0 & 0 \\
                          {1\over{\sqrt{2}}} & {-1\over{\sqrt{2}}}
                          \end{array}
                  \right]
                  \Psi(n-1,t)  \;\;+\;\;
                  \left[  \begin{array}{cc}
                          {1\over{\sqrt{2}}} & {1\over{\sqrt{2}}} \\
                          0 & 0 
                          \end{array}
                  \right]
                  \Psi(n+1,t) \\
            & = & M_{+}  \Psi(n-1,t) + M_{-} \Psi(n+1,t),
\end{eqnarray*}
for matrices~$M_{+},M_{-}$ defined appropriately.
Note that this transformation is {\em unitary} on the basis states given
by~$\zee \times \zee_2$, since it is
the composition of a unitary operation and a reversible move to the left
or the right. Moreover,
since the particle starts at the origin with chirality state ``left'' (say), we have the initial conditions, $\Psi(0,0) = \left( \begin{array}{c} 1 \\ 0 \end{array} \right)$,
and~$\Psi(n,0) = \left( \begin{array}{c} 0 \\ 0 \end{array}\right)$ if~$n \not= 0$.

With the above formulation, the analysis of the Hadamard walk reduces to
solving a two dimensional linear recurrence (or a {\em difference
equation}). We show how this recurrence may be analysed in the next
section.

\subsection{Fourier analysis of the Hadamard walk}
\label{sec-fourier}

As mentioned in Section~\ref{sec-intro}, quantum walk of the kind
defined above has, due to translational invariance, a very simple description in the Fourier domain. We therefore cast the problem of time evolution in this basis, where it can be easily solved, and at the end revert back to the real space description by inverting the Fourier transformation. The details are described below.

The spatial Fourier transform~$\tpsi(k,t)$ (for~$k \in
[-\pi,\pi]$) of the wave function~$\Psi(n,t)$ over~$\zee$ 
is given by
(cf.\ Appendix~\ref{sec-tech})
\begin{eqnarray*}
\tpsi(k,t) & = & \sum_n \Psi(n,t) \, \ee^{\aai kn}.
\end{eqnarray*}
In particular, we have~$\tpsi(k,0) = \left(\begin{array}{c} 1 \\ 0
\end{array}\right)$ for all~$k$, for a
particle starting at the origin with initially ``left'' chirality.

From the dynamics of~$\Psi$, we may deduce the following about~$\tpsi$:
\begin{eqnarray*}
\tpsi(k,t+1)
    & = & \sum_n \left( M_{+} \Psi(n-1,t) + M_{-} \Psi(n+1,t) \right)
              \, \ee^{\aai kn} \\
    & = & \ee^{\aai k} M_{+} \sum_n \Psi(n-1,t) \, \ee^{\aai k(n-1)}
          + \ee^{-\aai k} M_{-} \sum_n \Psi(n+1,t) \, \ee^{\aai k(n+1)}
          \\
    & = & \left( \ee^{\aai k} M_{+} + \ee^{-\aai k} M_{-} \right) 
          \tpsi(k,t).
\end{eqnarray*}
Thus, we have, 
\begin{equation}
\label{eqn-mk}
\tpsi(k,t+1) = M_k \tpsi(k,t)
\end{equation}
where
\begin{eqnarray}
M_k &=& \ee^{\aai k}M_{+} + \ee^{-\aai k} M_{-} \\
    &=& {1\over{\sqrt{2}}}
                   \left[  \begin{array}{cc}
                          \ee^{-\aai k}  & \ee^{-\aai k} \\
                          \ee^{\aai k} & -\ee^{\aai k}
                          \end{array}
                  \right].
\end{eqnarray}

Note that~$M_k = \Lambda_k U^T$, where~$\Lambda_k$ is the diagonal
matrix with entries~$\ee^{-\aai k}, \ee^{\aai k}$
and~$U^T$ is the transpose of the unitary matrix~$U$ that acts on the chirality state of the particle. Hence~$M_k$ is a unitary matrix. [This general
presentation anticipates the walk with general U.]

The recurrence in Fourier space thus takes the simple form~$\tpsi(k,t+1)
= M_k \, \tpsi(k,t)$, leading to~$\tpsi(k,t) =  M_k^t \, \tpsi(k,0)$. 
We may calculate~$M_k^t$ (and thus~$\tpsi(k,t)$)
by diagonalising the matrix~$M_k$, which is readily done
since it is a~$2\times 2$ unitary matrix.
If~$M_k$ has eigenvectors ($\ket{\Phi^1_k}$,$\ket{\Phi^2_k}$)
and corresponding eigenvalues ($\lambda^1_k$,$\lambda^2_k$),
we can write:
$$ 
M_k = \lambda^1_k \, \density{\Phi^1_k} \;\;+\;\;
      \lambda^2_k \, \density{\Phi^2_k},
$$
and then immediately we obtain the time evolution matrix as:
\begin{eqnarray*}
M_k^t & = &
    (\lambda^1_k)^t \, \density{\Phi^1_k}
    \;\;+\;\; (\lambda^2_k)^t \, \density{\Phi^2_k}.
\end{eqnarray*}
The eigenvalues of~$M_k$ are~$\lambda^1_k
= \ee^{-\aai \omega_k}$ and~$\lambda^2_k
= \ee^{\aai (\pi + \omega_k)}$,
where~$\omega_k$ is defined as the angle
in~$[-{\pi\over 2},{\pi\over 2}]$
such that~$\sin (\omega_k) = {{\sin k}\over\sqrt{2}}$.
The corresponding eigenvectors are displayed in Appendix~\ref{App1}.
For definiteness, here we consider the time evolution of a
particle that begins at the origin in the ``left'' chirality state. In 
the Fourier basis this initial state is represented by~$\tpsi(k,0) =
(1,0)^T$ for all~$k$. The wave function at time~$t$ is then given by:
\begin{eqnarray}
\tilde{\psi}_\lft(k,t) & = &
    \frac{1}{2}
    (1+\frac{\cos k}{\sqrt{1+\cos^2 k}}) \,
    \ee^{-\aai \omega_k t}
    \;\;+\;\;
    \frac{(-1)^t}{2}
    (1-\frac{\cos k}{\sqrt{1+\cos^2 k}}) \,
    \ee^{\aai \omega_k t} \\
\tilde{\psi}_\rght(k,t) & = &
     \frac{\aai \ee^{\aai k}}{2\sqrt{1+\cos^2 k}}
     (\ee^{-\aai \omega_k t}-(-1)^t\ee^{\aai \omega_k t})
\end{eqnarray} 

We now invert the Fourier transformation, to return to the basis in real space
(cf.\ Appendix~\ref{sec-tech}).
The wave functions in real space can be written in the form:

\begin{eqnarray}
\psi_\lft(n,t) & = & \frac{1+(-1)^{n+t}}{2}
                  \int_{-\pi}^{\pi} \, \frac{dk}{2\pi} \,
                  (1+\frac{\cos k}{\sqrt{1+\cos^2 k}}) \,
                  \ee^{-\aai (\omega_k t + kn)} \\
\psi_\rght(n,t) & = & \frac{1+(-1)^{n+t}}{2}
                  \int_{-\pi}^{\pi} \, \frac{dk}{2\pi} \,
                  \frac{\ee^{\aai k}}{\sqrt{1+\cos^2 k}} \,
                  \ee^{-\aai (\omega_k t + kn)}
\end{eqnarray}

Notice that the amplitudes vanish for even~$n$ (respectively, odd~$n$)
at odd~$t$ (even~$t$), as we would expect from the definition of the walk.

Thus, we have obtained a closed form solution for the time
evolved wave function of the Hadamard walk. In view 
of possible applications to developing Quantum algorithms, we are 
naturally led to considering the behaviour of the wave functions at 
large times ($t \gg 1$). Happily, the problem is considerably simplified 
in this asymptotic limit which allows us to accurately derive
several useful results. In the next section, we give details of this
asymptotic analysis.

\section{Asymptotic properties of the wave function in the large time limit}
\label{sec-asympt}

In the previous section, we obtained an exact solution for the time
evolution of the Hadamard walk.
In what follows, we will use
extensively the Method of Stationary Phase (see Appendix~\ref{sec-tech})
to extract the asymptotic properties of the resulting wave function.

\subsection{The asymptotic probability distribution}

\begin{figure}[t]
\begin{center}
\epsfxsize=3.2in
\hspace{0in}
\epsfbox{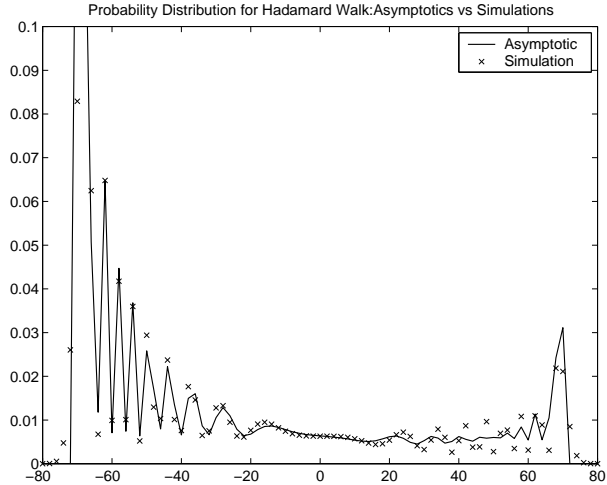}
\caption{\it A comparison of two probability distributions, one obtained
from a computer simulation of the Hadamard walk, and the other from 
from an asymptotic analysis of the walk.
The number of steps in the walk was taken to be~$100$. Only the
probability at the {\em even\/} points is plotted, since the odd points
have probability zero.}
\label{fig-prob}
\end{center}
\end{figure}

The asymptotic 
analysis for~$\psi_\lft$ and~$\psi_\rght$ is essentially the
same. They can both be written as a sum of integrals of the
type~$I(\alpha,t)$ described in Appendix~\ref{App1}.
Here, $\alpha$ should be understood
as~$n/t$. Appendix~\ref{App1} gives the details of the analysis for this
generic integral; we summarise the conclusions below.

The asymptotics for the wave function are simple to describe. 
The wave function is essentially
uniformly spread over the interval between~$\mp 1/\sqrt{2}$ (and so there
its gross behaviour is like~$1/\sqrt{t})$, and it dies
out very fast (faster than any inverse polynomial in~$t$) 
outside the interval. At the ``frontiers'' at~$\mp
1/\sqrt{2}$, however, there are two peaks of width~$O(t^{1/3})$ where
the wave function goes as~$t^{-1/3}$. The net probability below these
two peaks and beyond~$\mp 1/\sqrt{2}$ thus goes as~$t^{-1/3}$, and does
not affect the properties of the wave function we are interested in. We
will therefore restrict our attention to the interval~$\alpha \in
(-1/\sqrt{2}, 1/\sqrt{2})$. The exact expressions for~$\Psi =
(\psi_\lft,\psi_\rght)$ for such~$\alpha$
are displayed in Appendix~\ref{App1}.

We can calculate the probability of observing the particle doing the
quantum walk at any given point~$n = \alpha t$ from the wave function
derived above. Below we give the asymptotic
distribution for points~$\alpha = n/t$
between~$-1/\sqrt{2} + \epsilon$ and~$1/\sqrt{2} - \epsilon$, for an
arbitrarily small constant~$\epsilon > 0$.
\begin{eqnarray}
P(\alpha, t)
    & = & \size{\psi_\lft(\alpha t,t)}^2 +
                       \size{\psi_\rght(\alpha t,t)}^2 \nonumber \\
    & \sim & \frac{1+(-1)^{(\alpha+1)t}}
             {\pi t \,|{\omega''_{k_\alpha}}|}
             \times \nonumber \\
\label{eqn-prob}
    &      & \left[ (1-\alpha)^2 \cos^2 (\phi(\alpha) \, t
                    + \pi/4) \;\;+\;\; (1-\alpha^2) \cos^2 (
                    \phi(\alpha) \, t + k_\alpha + \pi/4)
             \right],
\end{eqnarray}
where~$\phi(\alpha) = -\omega_{k_\alpha} - \alpha k_\alpha$,
and~$k_\alpha$ is the
root of~$\omega'_k + \alpha = 0$ in~$[0,\pi]$.

The (approximate)
probability distribution~$P$ compares very well with simulation
results even for small~$t$,
as is evident from Figure~\ref{fig-prob}. The bias to the left
in the probability distribution plotted in the figure
is an artifact of the choice of initial chirality state (it was chosen
to be ``left''). If the particle begins in the chirality
state~$\frac{1}{\sqrt{2}}(\ket{\lft} + \aai\ket{\rght})$,
the distribution at any time
turns out to be symmetric (cf.\ Section~\ref{sec-general}, and
Figure~\ref{fig-symmprob} in the appendix). Indeed, the
Hadamard walk is an {\em unbiased\/} walk.

\subsection{Calculation of the moments}

We show in Appendix~\ref{App2} that the net probability of the
points~$n$ with~$\alpha = n/t$ between~$-1/\sqrt{2} + \epsilon$
and~$1/\sqrt{2} - \epsilon$, where~$\epsilon$ is an arbitrarily small
constant, is~$1 - \frac{2\epsilon}{\pi} - \frac{O(1)}{t}$, so the rest of
the points do not contribute to any global properties of the
distribution. Henceforth, we restrict ourselves to this interval.

\begin{figure}
\begin{center}
\epsfxsize=3.2in
\hspace{0in}
\epsfbox{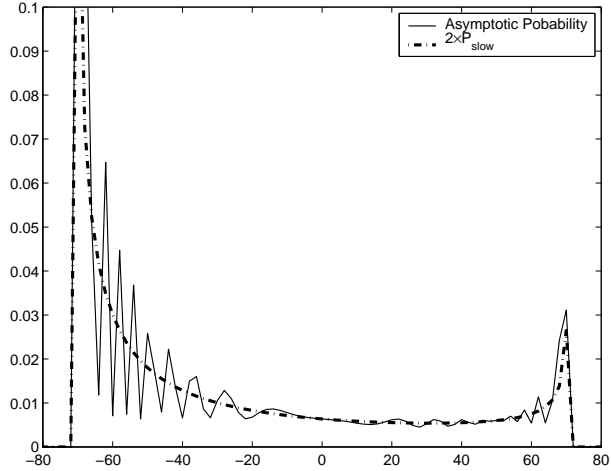}
\caption{\it A comparison of the distributions~$P$ and~$P_\slow$ for~$t
= 100$.
Only the probability at the {\em even\/} points is plotted,
and~$P_\slow$ is scaled by a factor of~$2$ because it has support on the
odd points as well.
}
\label{fig-pslow}
\end{center}
\end{figure}

For the purposes of calculating the moments of the distribution, it
will be convenient to decompose~$P$ as
\begin{equation}
P(\alpha, t) \;\;=\;\; P_\slow(\alpha, t) + P_\fast(\alpha,t),
\end{equation}
where
\begin{eqnarray}
P_\slow(\alpha, t)
    & = & \frac{1-\alpha}{\pi t \,|{\omega''_{k_\alpha}}|} 
\end{eqnarray}
is a slowly varying (non-oscillating) function in~$\alpha$, and~$P_\fast$
is the remaining (quickly oscillating) component. By the
analysis shown in Appendix~\ref{App2}, we can
show that any contribution to a moment from the
``fast'' component~$P_\fast$ is of lower order in~$t$ than the contribution
from~$P_\slow$. In Figure~\ref{fig-pslow}, we
compare~$P_\slow$ with~$P$.

The calculation of moments is further
simplified by the following observation.
Let~$p(\alpha) = t P_\slow(\alpha,t)$.
Then~$p$ is a probability
density function over the interval between~$\mp 1/\sqrt{2}$. It is
clearly non-negative, and we show below that it integrates to~$1$.
This observation allows us to approximate the sums in the moment
calculations by a {\em Riemann integral}, and the error so introduced is
again a lower order term (cf.\ Appendix~\ref{App2}).

To see that~$p(\alpha)$ integrates to~$1$, note
that~$\frac{\partial \phi}{\partial k}(k_\alpha, \alpha) =
0 = -\omega'_{k_\alpha} - \alpha$, so
\begin{equation}
|\omega''_{k_\alpha}| \;\;\equiv\;\; -\omega''_{k_\alpha} \;\;=\;\;
\frac{d\alpha}{dk_\alpha}.
\end{equation}
Now,
$$
\int_{-1/\sqrt{2}}^{1/\sqrt{2}} \, p(\alpha)\, d\alpha 
    \;\; = \;\;   \frac{1}{\pi} \int_{-1/\sqrt{2}}^{1/\sqrt{2}} \, (1-\alpha) 
                  \frac{dk_\alpha}{d\alpha} \, d\alpha
    \;\; = \;\;   \frac{1}{\pi} \int_{0}^{\pi} \, \left( 1 + \frac{\cos
                  k}{\sqrt{1+\cos^2 k}} \right) dk
    \;\; = \;\;   1,
$$
since the function~$\cos k/\sqrt{1+\cos^2 k}$ is anti-symmetric around
the point~$\pi/2$.

The different moments for the density function~$p$
are now readily
calculated by standard methods from complex residue theory.
These are listed in
Figure~\ref{fig-moments} for comparison with simulation
results.\footnote{As mentioned before, the particle has a constant speed
to the left, as indicated by its mean position, which is a result of its
biased initial state. For an unbiased initial state, the mean would be
zero.}
This gives us the leading term for the moments for the distribution~$P$.

\begin{figure}
\begin{center}
\begin{tabular}{|l|c|c|}
\hline
     & $p(\alpha)$ & simulation \\
\hline
 & & \\  
$\langle \alpha \rangle$ & $-1 + 1/{\sqrt{2}} = -0.293$
    & $-0.293$ \\
 & & \\  
$\langle\size{\alpha}\rangle$ & $1/ 2$ & $0.500$ \\
 & & \\  
$\langle \alpha^2 \rangle$ & $1- {1/{\sqrt{2}}} = 0.293$ & $0.293$ \\
\hline
\end{tabular}
\caption{\it A table of moments calculated with the approximation by the
density function~$p(\alpha)$, which are compared with computer
simulation results with~$t = 80$.}
\label{fig-moments}
\end{center}
\end{figure}

\subsection{Mixing behaviour of the walk}

It is quite evident from Figure~\ref{fig-prob} that the probability
distribution~$P$ is almost uniform over the interval between~$-1/\sqrt{2},
1/\sqrt{2}$. We argue this formally below.

Recall that the~$\delta$-mixing time ($\tau_\delta$)
of a random process is defined as
the first time~$t$ such that the distribution at time~$t$ is at total 
variation distance (which is half the~$\ell_1$ distance) at
most~$\delta$ from the uniform distribution. We claim that there is a
constant~$\delta < 1$ such that at time~$t$, $P$ is~$\delta$-close to
the uniform distribution on the integer points between~$\mp t/\sqrt{2}$.
We emphasise that for classical symmetric random walk, the corresponding
mixing time is {\em quadratic\/} in~$t$.

In order to show that the Hadamard walk is ``mixed'' at time~$t$ (in the
sense described above), it suffices to show that a constant
fraction~$\beta > 0$ of the points in the said interval have probability
between~$c/\sqrt{2}\,t$ and~$1/\sqrt{2}\,t$
for a constant~$c > 0$. A straightforward
calculation shows that this implies that the~$\ell_1$ distance from
uniform over~$[-t/\sqrt{2},\, t/\sqrt{2}]$ is at most~$2(1-\beta c)$, so
that~$\delta = 1 - \beta c < 1$ is a constant.

As in the previous section,
we restrict ourselves to the interval where~$n/t = \alpha \in
[-1/\sqrt{2} + \epsilon, \; 1/\sqrt{2} - \epsilon]$ (with~$\epsilon$
chosen to be a suitable constant) such
that~$P(\alpha,t)$ is at most~$1/\sqrt{2} t$ within the interval.
Recall that the probability mass within this interval is~$1 -
\frac{2\epsilon}{\pi} - \frac{O(1)}{t}$, which is a constant greater
than~$0$ (see Appendix~\ref{App2} for a proof of this). Clearly, this
cannot hold unless at least a constant fraction~$\beta = 1 -
\frac{2\epsilon}{\pi} - c$ of the points within
this interval have probability at least some constant~$0 < c < 1 -
\frac{2\epsilon}{\pi}$ over~$\sqrt{2}\,t$.
This completes the proof of the mixing nature of
the Hadamard walk.

\section{Discussion}
\label{sec-concl}

In this paper, we defined and 
studied the behaviour of quantum walks on the line, and showed a linear
time mixing behaviour for these walks. This has an immediate
bearing on the similar walks defined on the circle. It is easy to derive
the wave function for the walk on the circle, since it is exactly the
wave function for the unbounded line wrapped around the circle. The
linear time mixing for the circle is immediate. It is also possible to
derive the wave function for an appropriately modified walk on the {\em
finite\/} line using the solution to the unbounded walk. We expect that
this walk be mixing in linear time 
as well, but leave the details to~\cite{NayakV00}.
(Note that because of boundary effects, the mixing of the walk on a
finite line does not follow directly from our results
if the walk begins close to the edges.)

Aharonov {\it et al.\/}~\cite{AharonovAKV00} have
defined quantum walks on general graphs in the same
manner. It would be interesting to characterise the class of graphs on
which quantum walks mix faster than classical ones. A promising
candidate is the class of Cayley graphs, which seem to be
amenable to analysis along the same lines as ours
because of their rich group-theoretic structure. It would also be
interesting to study how much speed-up is possible in general, and for
different classes of graphs. The lower bound
result of~\cite{AharonovAKV00} is a step in this direction.

Our work leads to a plethora of questions regarding the properties of
quantum walks. It is yet unclear which of these are interesting from an
algorithmic point of view. Any speed-up of a known classical algorithm
based on random walk (such as those mentioned in
Section~\ref{sec-intro}) seems to involve the analysis of quantum walk on
graphs much more complex than the line. However, we believe such
analysis is still tractable.

Finally, we ask if it is possible to design quantum algorithms to
generate certain desirable superpositions 
efficiently---those corresponding to
distributions that are classically hard to sample from
(much in the spirit of~\cite{NayakSV97}).
It is conceivable that such algorithms follow an approach that is
completely different from the standard Markov chain Monte Carlo
method.

\subsection*{Acknowledgements}

We would like to thank Mike Saks for discussions on the asymptotic
behaviour of integrals and for directing us to~\cite{BenderO78}.


\appendix

\section{Technical background}
\label{sec-tech}

In this section, we give details of the concepts and results from 
mathematical analysis that we use in the paper.

\subsection{The Fourier transform}

Let~$f : \zee \rightarrow \complex$
be a complex valued function over the integers. Then its
Fourier transform~$\teff : [-\pi,\pi] \rightarrow \complex$ is
defined as
$$
\teff(k) \;\;=\;\; \sum_n f(n) \, \ee^{\aai kn}.
$$
The corresponding inverse Fourier transform is then given by
$$
f(n) \;\;=\;\; \frac{1}{2\pi} \int_{-\pi}^{\pi} \teff(k) 
               \, \ee^{-\aai kn} \, dk.
$$
Note that this is the dual view of the Fourier transform on the space of
functions on~$[-\pi,\pi]$.

We will be concerned with functions~$f$ with finite support, i.e.,
functions which are zero except at
finitely many points~$n$. The Fourier transforms are extremely well behaved
for this class of functions.
For more details on the properties of these
transforms, see, for example,~\cite{DymM72}.

\subsection{Asymptotic expansion of integrals}

Studying the large time behaviour of quantum walks naturally leads us to
consider the behaviour of integrals of the form
\begin{eqnarray}
\label{eqn-integral}
I(t)  & = & \int_{a}^{b} g(k) \ee^{\aai \, t \, \phi(k)} \, dk
\end{eqnarray}
as~$t$ tends to infinity. There is a well-developed theory of the
asymptotic expansion of integrals which allows us to determine, very
precisely, the leading terms in the expansion of the integral in terms
of simple functions of~$t$ (such as inverse powers of~$t$).
Below, we explain an important technique known as
the method of stationary phase~\cite{BenderO78, BleisteinH75} from this
theory.

Intuitively, the method may be understood as follows.
The exponential in the integral is a rapidly oscillating
function if~$t$ is large (and if~$\phi$ is not constant in any
subinterval). If~$g$ is a smooth function of~$k$, then the contributions
from adjacent subintervals nearly cancel each other out, and the major
contribution to the value of the integral comes from the region where
the oscillations are least rapid.
The regions of slow oscillation occur
precisely at the {\em stationary points\/}
of the function~$\phi$, i.e, points~$c$
where~$\phi'(c) = 0$. (If no such point exists, we can get the
asymptotic expansion for~$I$ in terms of inverse powers of~$t$
by repeated integration by parts, and the integral decays faster
than~${1\over t}$.) So the significant terms in the expansion of the
integral come from a small interval around the stationary points.
The ``flatness'' of~$\phi$ at a stationary point then
determines the contribution of its neighbourhood to the integral.
For example, if~$\phi''(c) \not= 0$, then the integral goes as~${1 \over
{\sqrt{t}}}$, however if~$\phi''(c) = 0$ but~$\phi'''(c) \not= 0$, then the
integral goes as~${1\over {t^{1/3}}}$.

Without going into the details of its derivation, we state the leading term
in the expansion of~$I(t)$ assuming that 
it has exactly one stationary point,
and that it occurs at the left end point~$a$
of the interval. (Any integral may be written as a sum of such
integrals.) We also assume that~$g$ is smooth and non-vanishing at~$a$.
Suppose~$a$ is a stationary point of order~$p-1$, i.e.,
$\phi'(a) = \phi''(a) = \cdots = \phi^{(p-1)}(a) = 0$ but~$\phi^{(p)}(a)
\not= 0$, then the dominant behaviour of~$I$ is given by
\begin{equation}
\label{eqn-asympt}
I(t) \sim g(a) \, \ee^{\aai t \phi(a) \pm \aai \pi/2p}
               \left[ \frac{p!}{t\size{\phi^{(p)}(a)}} \right]^{1/p}
               \frac{\Gamma(1/p)}{p}, ~~~~t \rightarrow +\infty,
\end{equation}
where we use the factor~$\ee^{\aai \pi/2p}$ (respectively, $\ee^{-\aai
\pi/2p}$) if~$\phi^{(p)}(a) > 0$ (if~$\phi^{(p)}(a) < 0$).
For the reader interested in more details about this and other associated
techniques, we
recommend~\cite{BenderO78, BleisteinH75}.

\section{Solution to the Hadamard walk}
\label{App1}

The following are the two eigenvectors of the matrix~$M_k$ for the Hadamard
walk mentioned in Section~\ref{sec-fourier}. (Recall
that~$\omega_k$ is defined as the angle in~$[-{\pi\over 2},{\pi\over
2}]$ such that~$\sin (\omega_k) = {{\sin k}\over\sqrt{2}}$.)
\begin{eqnarray*}
\Phi^1_k
    & = & {1 \over\sqrt{2}}
          \left( (1+ \cos^2 k) - \cos k \sqrt{1 + \cos^2 k}
          \right)^{-{1\over 2}}
                       \left[
                         \begin{array}{c}
                           \ee^{-\aai k} \\
                           \sqrt{2}\, \ee^{-\aai \omega_k} - \ee^{-\aai k}
                         \end{array}
                       \right] \\
\Phi^2_k
    & = & {1 \over\sqrt{2}}
          \left( (1+ \cos^2 k) + \cos k \sqrt{1 + \cos^2 k}
          \right)^{-{1\over 2}}
                       \left[
                         \begin{array}{c}
                           \ee^{-\aai k} \\
                           -\sqrt{2}\, \ee^{\aai \omega_k} - \ee^{-\aai k}
                         \end{array}
                       \right].
\end{eqnarray*}
We now get the expression for the Fourier transform at time~$t$, given
the initial state from the identity
\begin{eqnarray*}
\tpsi(k,t) & = & \ee^{-\aai \omega_k t} \;
                 \braket{\Phi^1_k}{\tpsi(k,0)} \;
                 \Phi^1_k
                 \;\;+\;\;
                 \ee^{\aai (\pi + \omega_k) t} \;
                 \braket{\Phi^2_k}{\tpsi(k,0)} \;
                 \Phi^2_k.
\end{eqnarray*}
The expressions for~$\tilde{\psi}_\lft, \tilde{\psi}_\rght$
stated in Section~\ref{sec-fourier} are derived by
calculating this for a 
particle that begins the walk at the origin with chirality ``left''
(i.e., with~$\tpsi(k,0) = (1,0)^T$ for all~$k$).

We now turn to determining
the asymptotic behaviour of the wave function~$\Psi$ as~$t$ tends
to~$+\infty$. It suffices to consider an integral of the form
\begin{eqnarray}
I(\alpha,t) & = & \int_{-\pi}^{\pi} \, \frac{dk}{2\pi} \,
                  g(k) \, \ee^{\aai \, \phi(k,\alpha) \, t},
\end{eqnarray}
where~$g(k)$ is analytic, and a periodic function of period~$2\pi$ taken to be
either even or odd, $\phi(k,\alpha) = -(\omega_k + \alpha k)$,
and~$\alpha \in [-1,1]$. The integrals in the expressions
for~$\psi_\lft, \psi_\rght$ are exactly of this kind.

It turns out that for large~$t$,
the integral~$I(\alpha,t)$ has three distinct kinds of
behaviour depending on the value of~$\size{\alpha}$,
with sharp transitions
from one behaviour to the other. It decays faster than any inverse
polynomial in~$t$
for~$\size{\alpha} > {1\over\sqrt{2}}+ O(t^{-2/3})$, goes
as~$t^{-1/3}$ in an~$O(t^{-2/3})$
interval around~$\pm \frac{1}{\sqrt{2}}$, and
as~$t^{-1/2}$ in the remaining interval.
Below, we show these behaviours for~$I$  for~$\size{\alpha} >
{1\over\sqrt{2}}+\epsilon$, $\size{\alpha} = \frac{1}{\sqrt{2}}$,
and~$\size{\alpha} < {1\over\sqrt{2}} - \epsilon$, for any
constant~$\epsilon > 0$, and leave the more detailed analysis of the
transitions from one behaviour to another to~\cite{NayakV00}.

First, we concentrate on~$\size{\alpha}$ larger than~${1\over\sqrt{2}}$
by a constant. For this range of~$\alpha$, $\phi$ does not have any
stationary points, and we can use integration by parts to show the
vanishing nature of~$I$.

Before we sketch a proof for this, we calculate some quantities that
will also be of use later. 
\begin{eqnarray}
\frac{\partial \phi}{\partial k} & = &  -(\omega'_k + \alpha) \;\;=\;\;
    - \frac{\cos k}{\sqrt{1 + \cos^2 k}} - \alpha \\
\frac{\partial^2 \phi}{\partial k^2} & = & -\omega''_k \;\;=\;\;
    \frac{\sin k}{(1+\cos^2 k)^{3/2}} \\
\frac{\partial^3 \phi}{\partial k^3} & = & -\omega'''_k \;\;=\;\;
    \frac{2 \cos k \, (1+\sin^2 k)}{(1+\cos^2 k)^{5/2}}
\end{eqnarray}

Note that for~$\size{\alpha} \ge 1/\sqrt{2} + \epsilon$,
$\size{\frac{\partial \phi}{\partial k}} \ge \epsilon$, so
its inverse is an analytic function in~$k$. Let
$$
u(k) \;\; = \;\; \frac{g(k)}{\aai\, t\, \frac{\partial \phi}{\partial k}},
~~~~{\mathrm{and}}~~~~
v(k) \;\; = \;\; \ee^{\aai \phi(k) \, t}.
$$
(In the above, we suppress the dependence of the functions
on~$\alpha$ for ease of notation.)
Now, integrating by parts,
$$
I(\alpha,t) \;\; \equiv \;\; 
        \int_{-\pi}^{\pi} \, \frac{dk}{2\pi} \, u \,
        \frac{\partial v}{\partial k} 
    \;\;=\;\;
        [u(\pi) v(\pi) - u(-\pi) v(-\pi) ] - 
        \int_{-\pi}^{\pi} \, \frac{dk}{2\pi} \, v \,
        \frac{\partial u}{\partial k}.
$$
The first term above is~$0$ because of
periodicity, and the integrand in the second term is bounded (in
magnitude)
by~$c_{\epsilon}/t$ for some constant~$c_{\epsilon}$ 
that depends only on~$\epsilon$. To prove a similar statement for any
greater power of~$t$, we may use induction with the invariant that
the integral resulting from the previous integration by parts
is of the same form as~$I$.

We now look at the points~$\alpha = 1/\sqrt{2}, -1/\sqrt{2}$.
At these points, $\phi$ has a stationary point of order~$2$ at~$k =
\pi,0$, respectively, as may readily be verified.
Using the method of stationary phase, we thus get the following leading
term for~$I$ at these points:
\begin{eqnarray*}
I({1\over\sqrt{2}},t) & \sim &
    \frac{g(\pi)}{3\pi} \sqrt{2} \, \Gamma(1/3) 
    \left[ {6 \over t} \right]^{1/3}
    \cos\left( \frac{\pi}{\sqrt{2}} t + {\pi \over 6} \right) \\
I(-{1\over\sqrt{2}},t) & \sim &
    \frac{g(0)}{6\pi} \sqrt{{3\over 2}} \, \Gamma(1/3)
    \left[ {6 \over t} \right]^{1/3}.
\end{eqnarray*}

Finally, we turn to the interval of most interest to us,
$[-1/\sqrt{2}+\epsilon, 1/\sqrt{2}-\epsilon]$. When~$\alpha$ lies in
this region, $\phi$ has two stationary points~$k_\alpha, -k_\alpha$,
where~$k_\alpha \in [0,\pi]$ and
\begin{equation}
\cos k_\alpha = \frac{-\alpha}{\sqrt{1-\alpha^2}}.
\end{equation}
The phase and its second derivative at the stationary points are:
\begin{eqnarray}
\phi(\pm k_\alpha,\alpha) & = & 
    \mp (\omega_{k_\alpha} + \alpha k_\alpha) \\
\frac{\partial^2 \phi}{\partial k^2}(\pm k_\alpha,\alpha) \;\; = \;\;
    \mp \omega''_k  & = &  \pm (1-\alpha^2)\sqrt{1-2\alpha^2}.
\end{eqnarray}
We abbreviate~$\phi(k_\alpha,\alpha)$ by~$\phi(\alpha)$.
We can again employ the method of stationary phase to get the dominant
term in the expansion for~$I$:
\begin{equation}
I(\alpha,t) \;\;\sim\;\;
              \frac{g(k_\alpha)}{\sqrt{2\pi t \,|{\omega''_{k_\alpha}}|}}
              \times
              \left\{
                  \begin{array}{ll}
                      2 \cos(\phi(\alpha)\, t + \pi/4) 
                          & \mathrm{if}~g~\mathrm{is~even} \\
                      \ & \ \\
                      2 \aai \sin(\phi(\alpha)\, t + \pi/4)
                          & \mathrm{if}~g~\mathrm{is~odd}
                  \end{array}
              \right.
\end{equation}

Using the above analysis, we can write out the 
asymptotic expressions for~$\psi_\lft,\psi_\rght$:
\begin{eqnarray}
\left. \begin{array}{l}
       \psi_\lft(\alpha t, t) \\
       \ \\
       \psi_\rght(\alpha t, t)
       \end{array}
\right\} & \sim &
    \frac{1+(-1)^{(\alpha+1)t}}{\sqrt{2\pi t \,|{\omega''_{k_\alpha}}|}}
    \times
    \left\{ \begin{array}{l}
           (1-\alpha) \cos(\phi(\alpha)\, t + \pi/4) \\
           \ \\
           \alpha \, \cos(\phi(\alpha)\, t + \pi/4) \\
           - \sqrt{1-2\alpha^2} \,
           \sin(\phi(\alpha)\, t + \pi/4)
           \end{array}
    \right.
\end{eqnarray}
It is now straightforward to calculate the probability 
distribution~$P(\alpha, t)$
induced on the line by observing the position of the particle after~$t$
steps of the walk. Equation~\ref{eqn-prob} in Section~\ref{sec-asympt}
gives the asymptotic expression for~$P(\alpha, t)$.

\begin{figure}
\begin{center}
\epsfxsize=3.2in
\hspace{0in}
\epsfbox{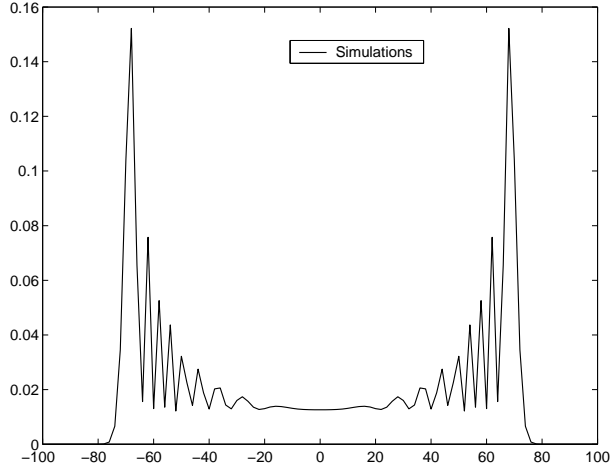}
\caption{\it The probability distribution obtained
from a computer simulation of the Hadamard walk with a symmetric initial
condition.
The number of steps in the walk was taken to be~$100$. Only the
probability at the {\em even\/} points is plotted, since the odd points
have probability zero.}
\label{fig-symmprob}
\end{center}
\end{figure}

Our estimate of the probability distribution is compared with computer
simulations in Figure~\ref{fig-prob} of Section~\ref{sec-asympt}. In
Figure~\ref{fig-symmprob}, 
we give a plot of the distribution obtained with the symmetric 
initial chirality state~$\frac{1}{\sqrt{2}}(\ket{\lft} +
\aai\ket{\rght})$.  The figure illustrates the unbiased
nature of the Hadamard walk.
We leave the details of the calculation of the asymptotic behaviour
of the distribution to~\cite{NayakV00}.

\section{Justification for approximations made in
         section~\ref{sec-asympt}}
\label{App2}

In calculating the moments of the probability distribution~$P(\alpha,t)$
obtained by observing a particle doing the Hadamard walk, we made two
approximations. First, we decomposed~$P$ into a slowly varying,
non-oscillating part~$P_\slow$ and a quickly oscillating part~$P_\fast$.
We claimed that the contribution to the moments from~$P_\fast$ is of
smaller order in~$t$. Then, we approximated the moment sums
involving~$P_\slow$ by a Reimann integral. In this section, we will
sketch proofs that the error involved in the two approximations both
go as~$1/t$ times the quantity being estimated.

As an important consequence of the error bound we derive, we
show that the net probability of points in the range~$\alpha
\in [-1/\sqrt{2} + \epsilon,\; 1/\sqrt{2} - \epsilon]$ differs from the
integral of the density function
\begin{eqnarray}
p(\alpha) & = & \frac{1-\alpha}{\pi \, |\omega_{k_\alpha}| }
\end{eqnarray}
over this interval by~$O(1)/t$. This integral
equals~$1 - \frac{2\epsilon}{\pi}$.

In the following discussion, we assume that~$n/t, \alpha
\in [-1/\sqrt{2} + \epsilon,\; 1/\sqrt{2} - \epsilon]$ (this range also
applies to sums over~$n$, integrals over~$\alpha$ etc.).

First, we bound the error in approximating a moment sum using~$P_\slow$
by an integral.
We compute the moments in terms of fractions of the appropriate powers
of~$t$. For the~$m$th moment ($m \ge 0$),
\begin{eqnarray}
\size{\sum_n  (n/t)^m P_\slow(n/t,t) -
        \int_{\alpha} \alpha^m p(\alpha,t)\, d\alpha}
    & \le & \sum_n \int_{n/t}^{(n+1)/t} \,d\alpha \,
            \size{(n/t)^m t\,P_\slow(n/t,t) - \alpha^m p(\alpha)}
            \nonumber \\
\label{eqn-error}
    & \le & \sum_n \int_{n/t}^{(n+1)/t} \,d\alpha \,
            \Delta_{1/t}(\alpha^m p(\alpha))
\end{eqnarray}
where
\begin{eqnarray*}
\Delta_{\delta} f(\alpha) & = & \max_\alpha \size{f(\alpha+\delta) -
f(\alpha)}.
\end{eqnarray*}
It is easy to verify that the variation given by~$\Delta_{1/t}$
for~$\alpha^m p(\alpha)$ over~$\alpha$ in the interval above
is~$O(1)/t$. This gives an error bound of~$O(1)/t$ in
equation~(\ref{eqn-error}) above.

We now show a bound of~$O(1)/t$ on the error introduced by the 
first kind of approximation mentioned above, namely by
ignoring~$P_\fast$, the component of~$P$
with quickly oscillating factors. We bound
one of the terms occurring in it; the others may be bounded similarly.
Consider the following term obtained from equation~(\ref{eqn-prob}) in
Section~\ref{sec-asympt}, by expressing the sine and cosine functions in
terms of exponentials:
\begin{eqnarray*}
\tpee(\alpha,t) & = & \frac{\alpha(1-2\alpha)}
                           {\pi t |\omega_{k_\alpha}|}
                      \ee^{2\aai \phi(\alpha)\, t},
\end{eqnarray*}
where~$\phi(\alpha) \eqdef \phi(k_\alpha,\alpha)$.
We estimate it's contribution to the~$m$th moment. The basic observation
is that each term in the moment sum such as~$\alpha^m \tpee(\alpha,t)$ 
is essentially
the leading term in the asymptotic expansion of an integral of the
type in equation~(\ref{eqn-integral}) 
discussed Appendix~\ref{sec-tech}:
\begin{eqnarray*}
\alpha^m \tpee(\alpha,t)
    & \sim & {1\over \sqrt{t}} \int_{\delta}^{\pi-\delta} \, dk \,
                               f(k)\, \ee^{2\aai \phi(k,\alpha)\, t},
\end{eqnarray*}                                             
where
\begin{eqnarray*}
 f(k) & = & -\frac{2 \omega'_k (1+2\omega'_k)}
            {\pi\sqrt{\pi |\omega''_k| }} \ee^{-\aai \pi/4},
\end{eqnarray*}
and~$\delta$ is a suitable constant.
The expression for~$f$ above
is derived using the identity~$\alpha = -\omega'_{k_\alpha}$ and
comparing the term in the moment sum with the asymptotic form of the
integral in equation~(\ref{eqn-asympt}) in Appendix~\ref{sec-tech}.
The error introduced by this integral representation is~$O(t^{-3/2})$.

We may now bound the contribution to the moment sum due to~$\tpee$ as
\begin{eqnarray*}
\sum_n (n/t)^m \tpee(\alpha,t)
    & \sim & {1\over \sqrt{t}} \sum_n \int_{\delta}^{\pi-\delta} \, dk
             \, f(k)\, \ee^{-2 \aai \omega_k t - 2\aai nk} \\
    &  =   & {1\over \sqrt{t}} \int_{\delta}^{\pi-\delta} \, dk
             \, f(k)\, \ee^{-2 \aai \omega_k t} 
             \sum_n \ee^{- 2\aai nk} \\
    &  =   & {1\over \sqrt{t}} \int_{\delta}^{\pi-\delta} \, dk
             \, f(k)\, \ee^{-2 \aai \omega_k t} \;
             \frac{\ee^{-2\aai k(\beta t + 1)} - \ee^{2\aai k\beta t}}
                  {\ee^{-2\aai k} - 1},
\end{eqnarray*}
where~$\beta = (1/\sqrt{2} -\epsilon)$. We may now use the method of
stationary phase again to derive the asymptotic form of the integral.
As we expect from the analysis we did in Appendix~\ref{App1}, the
integral goes as~$1/\sqrt{t}$, and thus the entire expression goes
as~$1/t$.

\section{General walk on the line}
\label{sec-general}

So far we have been describing the details of the 
Hadamard walk on the line. Here, we consider 
the properties of the general quantum walk on the line.
As mentioned in Section~\ref{sec-intro}, it suffices to
consider a particle that carries a
``left'' or ``right'' chirality, and 
at each time step undergoes a unitary transformation of its 
chirality state and then moves accordingly, one step to 
the right if it ends in ``right'' chirality 
state, and one step to the left if it ends in 
the ``left'' chirality state. The most general 
such unitary transformation contains four parameters. However, 
it can be shown~\cite{NayakV00}
that if we focus on the probability distribution that results 
from these walks, then there is a {\em single\/}
parameter that labels a family of walks, and every 
element of the family gives rise to identical probability 
distributions~$P(n,t)$ for the same initial conditions.
The walks given by this parameter are also singled out for study
in~\cite{Meyer96}.

We therefore consider a representative element from each family,
which we write as
\begin{eqnarray}
M(\theta) &=&  T \circ (I \tensor U_\theta), {\mathrm{~~~~where}} \\
T & = & T_- \tensor \density{\lft} + T_+ \tensor \density{\rght},
        {\mathrm{~~~~and}} \\
U_\theta &=& \ee^{\aai \, \frac{\theta}{2} \, \sigma_y}.
\end{eqnarray}
The operator~$M(\theta)$ evolves the state by one time
step, i.e., $\ket{\Psi(t+1)} = M(\theta) \ket{\Psi(t)}$,
the operator~$T$ implements the left/right shift ($T_+\ket{n}=\ket{n+1}$
while $T_-\ket{n}=\ket{n-1}$), and~$U_\theta$ is the unitary
transformation acting on the chirality state.
The Hadamard walk belongs to the 
family of walks parametrised by $\theta = \frac\pi2$. 
To show this we write the Hadamard transform as
\begin{equation}
H \;\; =\;\; \frac1{\sqrt{2}}\left( \begin{array}{cc}
			1 & 1\\ 
			1 & -1
		\end{array} \right)
  \;\; = \;\; -\aai \, \ee^{\aai\, \frac\pi2 \, \sigma_z} \;
              \ee^{\aai \, \frac\pi4 \, \sigma_y}
  \;\; = \;\; -\aai \, \ee^{\aai \, \frac\pi2 \, \sigma_z} \; U_{\frac\pi2}.
\end{equation}
Thus, by a suitable redefinition of phase of the chirality
states at time $t+1$, the additional rotation that 
multiplies~$U_{\frac\pi2}$ can be absorbed. Clearly, the
probability distributions for finding a particle at a point
are unaffected by this redefinition of phase.
Therefore, the Hadamard walk is 
equivalent to the walk with~$\theta=\frac{\pi}2$. 

The behaviour of the walk with~$\theta = 0$ and~$\theta = \pi$
are easily analysed. When~$\theta = 0$,
the unitary transformation rotating the chirality states is the
identity, and hence a particle starting with ``left'' 
(``right'') chirality, after $t$ time steps,
moves $t$ units to the left (right). This walk
is clearly not mixing, since the probability 
is concentrated at one or two points at any given instant,
and resembles more the biased classical walk
(though we wish to emphasise here that all the 
quantum walks we consider are {\em unbiased}, a point 
which we return to below).
Now, when $\theta = \pi$, the unitary transformation flips the 
chirality at each time step. Hence a particle zig zags 
between the origin and a neighbouring point, and thus 
has a non-zero probability on a maximum of three points.
This walk too is clearly not mixing.    

For intermediate values of $\theta$ the behaviour is 
more involved, as our analysis of the Hadamard walk indicates.
In fact, the qualitative behaviour of the Hadamard walk 
is generic to all~$\theta$ excluding the singular 
values~$0,\pi$ that we discussed above.
All of these walks are mixing, and mix in linear time.

We can calculate very easily some 
properties of the general walk, following the analysis for the Hadamard
walk.
We first note that the eigenvalues of~$M_k(\theta)$
for the general walk given by 
$\theta$ are
\begin{eqnarray}
\label{eqn-eigen}
\lambda^1_k(\theta), \lambda^2_k(\theta) & = &
                         \ee^{ \pm \aai \,\omega_k(\theta)},
                         {\mathrm{ ~~~~where}}\\
\cos(\omega_k(\theta)) & = & \cos{\frac\theta2}\cos{k} 
\end{eqnarray} 
where~$\omega_k(\theta) \in[0,\pi]$.

It is possible to deduce the relevant component~$P_\slow$
of the asymptotic probability distribution~$P$ from the eigenvalues in
equation~(\ref{eqn-eigen}),
as is evident from the case of the Hadamard walk.
The distribution~$P_\slow$
beginning with symmetric initial conditions is:
\begin{equation}
P_\slow(\alpha,t) \;\;=\;\; \frac{1}{\pi t
                     \, \omega_{k_\alpha}''(\theta) }
\end{equation}
and has support on the interval~$\alpha \in (-\cos\frac{\theta}{2},\;
\cos\frac{\theta}{2})$.
Here~$k_\alpha$ is given by the usual stationary point condition:
\begin{equation}
\left. \frac{d\omega_k(\theta)}{dk} \right|_{k = k_\alpha} \;\;=\;\; \alpha,
\end{equation}
and the derivatives of~$\omega_k(\theta)$ with respect to~$k$ are
\begin{eqnarray}
\omega'_k(\theta) & = & \frac{\sin k \cos \frac{\theta}{2}}
                        {\sin(\omega_k(\theta))} \\
\omega''_k(\theta) & = & \cot(\omega_k(\theta))\, [ 1 -
                         (\omega'_k(\theta))^2 ].
\end{eqnarray}
Thus the stationary points are~$k_\alpha, \pi-k_\alpha$, where~$k_\alpha
\in [-\pi/2,\; \pi/2]$ and
\begin{eqnarray}
\sin k_\alpha & = & \frac{\alpha}{\sqrt{1-\alpha^2}} \,
                    \tan \frac{\theta}{2}.
\end{eqnarray}
The second derivative of the phase at the stationary points is~$\pm
\omega''_{k_\alpha}(\theta)$ where
\begin{eqnarray}
\omega''_{k_\alpha}(\theta) & = & \frac{1-\alpha^2}
                    {\sin\frac{\theta}{2}} 
                    \sqrt{\cos^2 \frac{\theta}{2} - \alpha^2}.
\end{eqnarray}

We can readily verify that the probability sums to one, i.e.,
\begin{equation}
t\int d\alpha \, P_\slow(\alpha,t) \;\;=\;\;
     \int d\alpha \, \frac1{\pi \, \omega_{k_\alpha}''(\theta)} 
     \;\;=\;\; \int_{-\pi/2}^{\pi/2} \, \frac{dk_\alpha}{\pi}
     \;\;=\;\; 1,
\end{equation}
where we have used the fact that $\omega''_{k_\alpha} =
\frac{d\alpha}{dk_\alpha}$. Thus, as in the case of the Hadamard walk,
almost all the probability is confined to the interval between~$\mp \cos
{\theta\over 2}$ and is close to uniform over it.
However, the ``height'' of the distribution is proportional to~$\sin
{\theta\over 2}$, and the distance from uniform increases as~$\theta
\rightarrow 0$. On the other hand,
as~$\theta \rightarrow \pi$, the ``width''
of the distribution
(which is proportional to~$2\cos {\theta\over 2}$)
goes to zero, again increasing the distance from uniform.

The distribution~$P_\slow$ can, as before, be used to calculate
moments. For example, the mean position of the
particle~$\langle|\alpha|\rangle$ is given by
\begin{equation}
\int d\alpha \, |\alpha| \, t P_\slow(\alpha,t) \;\; = \;\; 
    2\int_0^\frac\pi2 \frac{dk_\alpha}{\pi} \,
    \omega'_{k_\alpha}(\theta) \;\;=\;\; 1-\frac\theta\pi,
\end{equation}
which matches the expected value at the points $\theta=0,\frac\pi2,\pi$.

Finally we comment on the unbiased nature of the
quantum walks defined in this manner.
An unbiased walk (starting at the origin)
does not distinguish between the left and
right halves of the line, i.e., it is reflection symmetric.
In order to capture this into a condition on unbiased walks,
it is necessary to also account for the initial 
chirality state of the particle which can lead to a biased
probability distribution, although the rules governing the walk are 
intrinsically reflection symmetric. Therefore, although such a
symmetric walk can produce a biased probability distribution with some 
particular initial condition, there exists another initial condition 
for which the bias is exactly reversed. In more formal terms,
if we describe the quantum walk by the matrices $M_k$ in
equation~(\ref{eqn-mk}) in Section~\ref{sec-walk},
then we need to find a unitary matrix $S$ such that:
\begin{equation}
S^\dag M_k S = M_{-k}.
\end{equation}
The unitary matrix $S$ performs a rotation on the 
chirality state that reverses any bias arising from the 
choice of initial condition. If no such matrix exists,
then the walk intrinsically distinguishes left from right and is 
hence not an 
unbiased walk\footnote{Actually the condition is  a 
little more general, an overall sign change can also 
be permitted, i.e., $S^\dag M_k S = \pm M_{-k}$.
In fact, the minus sign appears for the Hadamard walk,
where $S=\sigma_y$.}.
An immediate consequence is that if we start with an 
initial chirality state that is an eigenvector of $S$,
then a symmetric probability distribution is generated. For the 
family of walks just considered, the unitary matrix
$S=\sigma_y$ will suffice 
since it only interchanges $T_+$ and $T_-$
and does nothing else.
Therefore, starting in the initial chirality 
state $\frac{1}{\sqrt{2}}(0, \;\pm \aai )^T$
generates a symmetric distribution,
as shown for the Hadamard walk in Figure~\ref{fig-symmprob}.


\begin{thebibliography}{10}

\bibitem{AharonovAKV00}
Dorit Aharonov, Andris Ambainis, Julia Kempe, and Umesh Vazirani.
\newblock Personal communication, October 2000.

\bibitem{AmbainisW00}
Andris Ambainis and John Watrous.
\newblock Personal communication, October 2000.

\bibitem{BenderO78}
Carl~M. Bender and Steven~A. Orszag.
\newblock {\em Advanced Mathematical Methods for Scientists and Engineers}.
\newblock International Series in Pure and Applied Mathematics. {McGraw-Hill,
  Inc.}, New York, 1978.

\bibitem{BleisteinH75}
Norman Bleistein and Richard~A. Handelsman.
\newblock {\em Asymptotic Expansions of Integrals}.
\newblock Holt, Rinehart and Winston, New York, 1975.

\bibitem{Diaconis88}
Persi Diaconis.
\newblock {\em Group Representations in Probability and Statistics}, volume~11
  of {\em Lecture Notes-Monograph Series}.
\newblock Institute of Mathematical Statistics, Hayward, California, 1988.

\bibitem{DyerFK91}
Martin Dyer, Alan Frieze, and Ravi Kannan.
\newblock A random polynomial-time algorithm for approximating the volume of
  convex bodies.
\newblock {\em Journal of the ACM}, 38(1):1--17, January 1991.

\bibitem{DymM72}
H.~Dym and H.~P. McKean.
\newblock {\em Fourier Series and Integrals}, volume~14 of {\em Probability and
  Mathematical Statistics}.
\newblock Academic Press, New York, 1972.

\bibitem{Grover96}
Lov~K. Grover.
\newblock A fast quantum mechanical algorithm for database search.
\newblock In {\em Proceedings of the Twenty-Eighth Annual {ACM} Symposium on
  the Theory of Computing}, pages 212--219, Philadelphia, Pennsylvania, 22--24
  May 1996.

\bibitem{JerrumS89}
Mark Jerrum and Alistair Sinclair.
\newblock Approximating the permanent.
\newblock {\em {SIAM} Journal on Computing}, 18(6):1149--1178, December 1989.

\bibitem{JerrumS96}
Mark Jerrum and Alistair Sinclair.
\newblock The {Markov} chain {Monte Carlo} method: An approach to approximate
  counting and integration.
\newblock In Dorit~S. Hochbaum, editor, {\em Approximation Algorithms for
  {NP}-hard Problems}, chapter~12, pages 482--520. {PWS} Publishing, Boston,
  1996.

\bibitem{JerrumSV00}
Mark Jerrum, Alistair Sinclair, and Eric Vigoda.
\newblock A polynomial-time approximation algorithm for the permanent of a
  matrix with non-negative entries.
\newblock Technical Report TR00-079, Electronic Colloquium on Computational
  Complexity, http://www.eccc.uni-trier.de/eccc/, 2000.

\bibitem{LovaszK99}
L\'aszl\'o Lov\'asz and Ravi Kannan.
\newblock Faster mixing via average conductance.
\newblock In {\em Proceedings of the Thirty-First Annual {ACM} Symposium on
  Theory of Computing}, Atlanta, Georgia, 1--4 May 1999. {ACM}.

\bibitem{Meyer96}
David~A. Meyer.
\newblock From quantum cellular automata to quantum lattice gases.
\newblock {\em Journal of Statistical Physics}, 85:551--574, 1996.

\bibitem{MotwaniR95}
Rajeev Motwani and Prabhakar Raghavan.
\newblock {\em Randomized Algorithms}.
\newblock Cambridge University Press, 1995.

\bibitem{NayakSV97}
Ashwin Nayak, Leonard Schulman, and Umesh Vazirani.
\newblock Unpublished, 1997.

\bibitem{NayakV00}
Ashwin Nayak and Ashvin Vishwanath.
\newblock In preparation, October 2000.

\bibitem{Papadimitriou91}
Christos~H. Papadimitriou.
\newblock On selecting a satisfying truth assignment.
\newblock In {\em Proceedings of the 32nd Annual Symposium on Foundations of
  Computer Science}, pages 163--169, San Juan, Puerto Rico, 1--4 October 1991.
  {IEEE} Computer Society.

\bibitem{RandallW99}
Dana Randall and David Wilson.
\newblock Sampling spin configurations of an ising system.
\newblock In {\em Proceedings of the Tenth Annual {ACM}-{SIAM} Symposium on
  Discrete Algorithms}, pages S959--960, Baltimore, Maryland, 17--19 January
  1999.

\bibitem{Schoning99}
Uwe Sch{\"o}ning.
\newblock A probabilistic algorithm for k-{SAT} and constraint satisfaction
  problems.
\newblock In {\em Proceedings of the 40th Annual Symposium on Foundations of
  Computer Science}, New York City, NY, 17--19 October 1999. {IEEE} Computer
  Society.

\bibitem{Shor97}
Peter~W. Shor.
\newblock Polynomial-time algorithms for prime factorization and discrete
  logarithms on a quantum computer.
\newblock {\em {SIAM} Journal on Computing}, 26(5):1484--1509, October 1997.

\bibitem{Simon97}
Daniel~R. Simon.
\newblock On the power of quantum computation.
\newblock {\em {SIAM} Journal on Computing}, 26(5):1474--1483, October 1997.

\bibitem{Watrous99a}
John Watrous.
\newblock Quantum simulations of classical random walks and gndirected graph
  connectivity.
\newblock In {\em Proceedings of the Annual {IEEE} Conference on Computational
  Complexity}, pages 180--187, Atlanta, Georgia, USA, 4--6 May 1999.

\end{thebibliography}
\end{document}